\documentclass{article}

\usepackage{arxiv}

\usepackage[utf8]{inputenc} 
\usepackage[T1]{fontenc}    
\usepackage{hyperref}       
\usepackage{url}            
\usepackage{booktabs}       
\usepackage{amsfonts}       
\usepackage{nicefrac}       
\usepackage{microtype}      
\usepackage{lipsum}
\usepackage{graphicx}
\graphicspath{ {./images/} }

\title{Concept Annotation for Intelligent Textbooks}

\author{
 Mengdi Wang*, Hung Chau*, Khushboo Thaker, Peter Brusilovsky, Daqing He\\
  School of Coumputing and Information\\
  University of Pittsburgh\\
  Pittsburgh, PA 15260 \\
  \texttt{\{mengdi.wang, hkc6, k.thaker,peterb,dah44\}@pitt.edu} 
  }

\begin{document}
\maketitle
\begin{abstract}
With the increased popularity of electronic textbooks, there is a growing interests in developing a new generation of  ``intelligent textbooks'', which have the ability to guide the readers according to their learning goals and current knowledge. The intelligent textbooks extend regular textbooks by integrating machine-manipulatable knowledge such as a knowledge map or a prerequisite-outcome relationship between sections, among which, 

the most popular integrated knowledge is a list of unique knowledge concepts associated with each section.
With the help of these concept, multiple intelligent operations, such as content linking, content recommendation or student modeling, can be performed. However, annotating a reliable set of concepts to a textbook section is a challenge. Automatic unsupervised methods for extracting key-phrases as the concepts are known to have insufficient accuracy. 
Manual annotation by experts is considered as a preferred approach and can be used to produce both the target outcome and the labeled data for training supervised models. However, most researchers in education domain still consider the concept annotation process  as an ad-hoc activity rather than an engineering task, resulting in low-quality annotated data. In this paper, we present a textbook knowledge engineering method to obtain reliable concept annotations. The approach has been applied to produce annotated concepts for \textit{Introduction to Information Retrieval} textbook. As shown by the data we collected, the inter-annotator agreement gradually increased along with our procedure, and the concept annotations we produced led to better results in document linking and student modeling tasks. The outcomes of our work include a validated knowledge engineering procedure, a code-book for technical concept annotation, and a set of concept annotations for the target textbook, which could be used as gold standard in further research.
\end{abstract}



\section{Introduction}

Modern textbooks have been developed and refined over many decades to evolve into well-organized tools for communicating knowledge and educating the next generation of professionals. Yet, the power of computing and internet caused the textbooks to evolve even faster than before. The conversion of textbooks into electronic format created an opportunity to augment textbooks with novel functionalities based on application or Artificial Intelligence. This direction of research, usually referred as ``intelligent textbooks'' explored a range of novel ideas over the last 20 years. The explored approaches include adaptive navigation support \cite{henze1999hyperbooks}, natural language question answering \cite{chaudhuri2013inquire}, automatic link creation\cite{guerra2013one}, and personalized recommendation of external content~\cite{agrawal2014study}.

The key to the power of most of the intelligent textbook technologies is ``knowledge behind pages'', which this technologies need to operate. These knowledge are usually extracted using a combination of machine learning, automatic natural language processing, and human knowledge engineering, i.e., annotation by human experts. Expert annotation is known to be of higher quality and is frequently used as the ``gold standard'' to assess the quality of automatic approaches. For some easier tasks such as content linking or content recommendation, automatic processing could support sufficient levels of quality. For more challenging tasks, such as personalization, the use of expert annotation in some form is essential. The problem is, however, that even an expert-level knowledge annotation might not achieve a quality required by intelligent approaches, unless it is guided by a reliable systematic procedure. In this paper we present our work on developing and evaluation of a systematic knowledge engineering approach for fine-grained annotation of textbooks with underlying knowledge in the form of concepts. Our study demonstrates that this approach produces better results in performance-based evaluation.

\section{Related Work}
\subsection{Intelligent Textbooks}
The research on intelligent textbooks could be traced back to the early attempts to develop electronic textbooks using pre-Web hypertext systems. At that time, artificial intelligence (AI) approaches were used to automate link creation between hypertext pages, which is an essential process to create a high quality hypertext~\cite{Bareiss1993applying}. Since these early attempts, ``intelligent linking'' remained as an integral part of hypertext research. A range of increasingly more advanced approaches to extract concepts and other semantic features from hypertext pages 
have been reported~\cite{green1999building,lakkaraju1999document,agrawal2014study,guerra2013one}.

The next generation of research on intelligent textbooks was motivated by the expanding World Wide Web and the migration of textbooks online. This generation focused on using adaptive hypermedia techniques to produce adaptive textbooks. By monitoring user reading and other activities in adaptive online textbooks, these systems attempted to model user knowledge and support the users with adaptive navigation within a book~\cite{henze1999hyperbooks,brusilovsky1998adaptive,kavcic2004fuzzy} as well as adaptive content presentation~\cite{melis2001activemath}. This generation of adaptive textbooks has been based on relatively advanced models of content annotation by domain experts, frequently using domain ontologies~\cite{brusilovsky2003developing}. Similar to automatic linking research, the research on concept-based adaptive textbooks remains active and focus on more advanced personalization technologies as well as automated domain model development and concept indexing.

The most recent generation of intelligent textbook was fueled by the increased availability of user data and focused on combining artificial and collective intelligence. Started with early attempts of using past users' behavior to provide social navigation support for future learners~\cite{brusilovsky2004social}, the research of this direction explored increasingly more complex approaches for mining past users' behavior to guide new users~\cite{lan2016} and predict their success~\cite{winchel2018}. Modern research on intelligent textbooks also frequently combines the ideas of automatic linking, personalization, concept annotation, and data mining~\cite{lan2016,Labutov2017}.

\subsection{Ground Truth Annotation}
Despite efforts to automate annotations of documents, manual annotations still play an important role in the construction of corpora for document engineering. The
quality of such manual annotations depends on a reliable coding schema. 
A coding schema can be seen as a set of guidelines to assign an objective (e.g., morphemes, words, phrases, sentences) to a single category. \cite{bayerl2003methodology} identified two considerations for a coding schema: 1) the categories of the coding schema must enable people to differentiate among the categories; and 2) the coding schema should be consistent among different coders or within one coder over different time. \cite{bayerl2003methodology} also proposed a methodological framework consisting of five successive steps for systematic schema development. Various schemata for ground truth annotation of documents were developed for different applications. For example, \cite{eryiugit2013turksent} explored sentiment annotation tools for sentiment analysis, which has gain high popularity and several academic projects emerged in this field. 
\cite{trani2014manual} proposed a manual annotation framework to link short fragments of text within a document for entity linking. \cite{berlanga2015tailored} used several knowledge bases for a semantic annotation strategy. 

\subsection{Concept Mining}
There are a wide range of applications related to concept mining such as \textit{key-phrase or concept extraction}, \textit{prerequisite-outcome concept prediction} \cite{Labutov2017}, or \textit{concept hierarchy creation} \cite{Wang2015}. Among these applications, concept extraction is the most fundamental task that leads to the success of other tasks; i.e., in order to predict a concept is a prerequisite or outcome concept we first need to identify if it is a concept.

Dozens of studies have tried to extract key-phrase automatically with different kinds of approaches including rules-based, supervised learning, unsupervised learning, and deep neural networks. However, their performance is still very low, making them are not effective enough to use for certain applications; for example, explainable recommendation systems. Typically, automatic key-phrase extraction systems consist of two parts. Firstly, they need to preprocess data and then extract a candidate keyphrase list with lexical patterns and heuristics \cite{Liu2009, Medelyan2009, Wang2015, Grineva2009, Mihalcea2004TextRank, Bougouin2013topic,LiuNAACL09, FlorescuACL17,Le2016}. Secondly, the candidates are ranked or classified to identify correct keyphrases using unsupervised methods or supervised with hand-crafted features. Candidates are scored based on some properties that show how likely a candidate being a keyphrase in the given document. Many studies have formed this task as a binary classification problem to determine correct keyphrases \cite{witten2005kea, Hulth2003, Jiang2009,Rose2010rake,Hulth2003,Wang2015}. 

For unsupervised learning, graph-based methods \cite{Mihalcea2004TextRank, Bougouin2013topic} try to find important keyphrases in a document. A candidate is important when it has relationships with other candidates and those candidates are also important in the document, forming a graph representing the input document, where a node and edge of the graph represents a keyphrase candidate and the relationship between two related candidates, respectively. Each node in the graph is assigned a score which can be calculated using ranking techniques such as \textit{PageRank}. Finally, they select the top-ranked candidates as keyphrases for the input document. On the other hand, topic-based clustering methods \cite{Liu2009,Liu2010,Grineva2009} group semantically similar candidates in a document as \textit{topics}. 
Keyphrases are then selected based on the centroid of each cluster or the importance of each topic.

Although deep neural networks have successfully applied to many NPL-related tasks, sequence tagging, named entity recognition, to name a few, few studies have focused on keyphrase extraction problem; and none of them have evaluated on textbook datasets. Meng et al.\cite{Meng2017CopyRNN} built a RNN-based  generative  model  using  encoder-decoder  architecture  to  predict  keyphrases. Though their performance was better than state-of-the-art methods, it was still not clear how to use in the educational setting since the datasets used to evaluated were scientific articles and paper abstracts. 

Wang et al. \cite{Wang2015} proposed a method for mining concept hierarchies for textbooks, which is also required to extract a list of concepts. In this study, instead of focusing concept extraction task, they use Wikipedia titles as a external resource to identify concepts appearing the textbook's table of content. As a result, there are only a few important extracted concepts considered as topic levels for building a hierarchy.

\section{Textbook Knowledge Annotation}

In education domain, knowledge annotation has been perform in many studies because its results often served as the primary input for the methods being developed. However, researchers usually perform it as an ad-hoc task and it is known to be a very challenging task. This is because 
it is hard to maintain consistency during the long process of annotation without clear rules and descriptions. 

To overcome this challenge, we designed a systematic textbook annotation procedure, and applied it in the annotation of a popular online available textbook \textit{Introduction to Information Retrieval} (IIR)  \footnote{https://nlp.stanford.edu/IR-book/}. The goal of our annotation is to add concepts to the book so that to turn it into an intelligent textbook, and this annotation task help us to refine the proposed textbook annotation procedure. 

\subsection{The Case Study: Introduction to Information Retrieval}
The ultimate goal of our research project is the development of intelligent textbooks, which could offer a rich set of support functionalities to their readers, including automatic linking and content recommendation. IIR textbook was one of our first targets. 
To support the expected functionalities, we have to produce a fine-grained annotation of concepts to this textbook. Before we introduce our systematic annotation approach, it is important to mention that in order to produce quality annotation for IIR textbook, we previously explored traditional ad-hoc expert annotation, crowdsourcing, concept extraction, and other approaches. While the overall quality of the obtained results and the inter-rater agreement for both experts and crowdworkers were lower than expected, the results of our earlier work were useful to guide our work on systematic annotation and to offer evaluation baselines.

\subsection{Initial Coding Procedure}

\begin{figure}[htbp]
\centering
    \includegraphics[width = 0.5\textwidth]{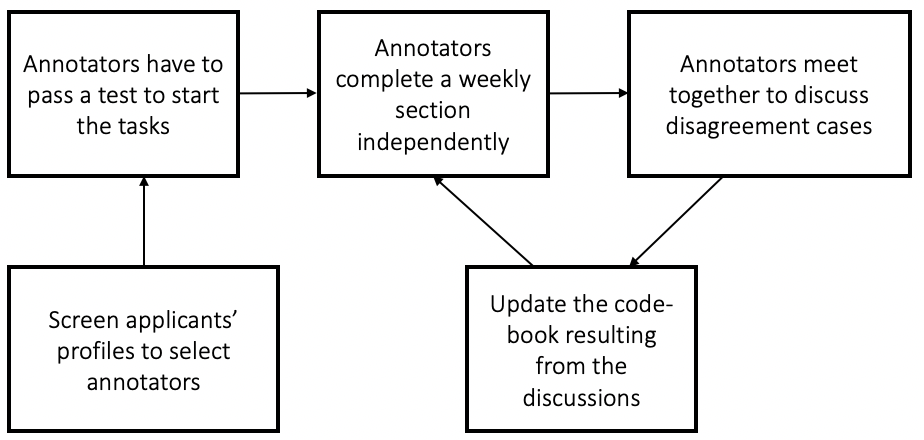}
    \caption{Coding procedure diagram. The annotators follow the procedure until they complete the whole process.}
    \label{fig:procedure_diagram}
\end{figure}

Our goal is to develop a systematic textbook annotation procedure so that high inter-annotator agreement can be achieved and maintained.
As shown in Figure \ref{fig:procedure_diagram}, the initial annotation procedure contains several standard steps including screening applicants' profiles, guiding annotators to perform the tasks and building an annotation code book. 

\subsection{Hiring Process} 
\begin{figure*}
    \includegraphics[width = 1\textwidth]{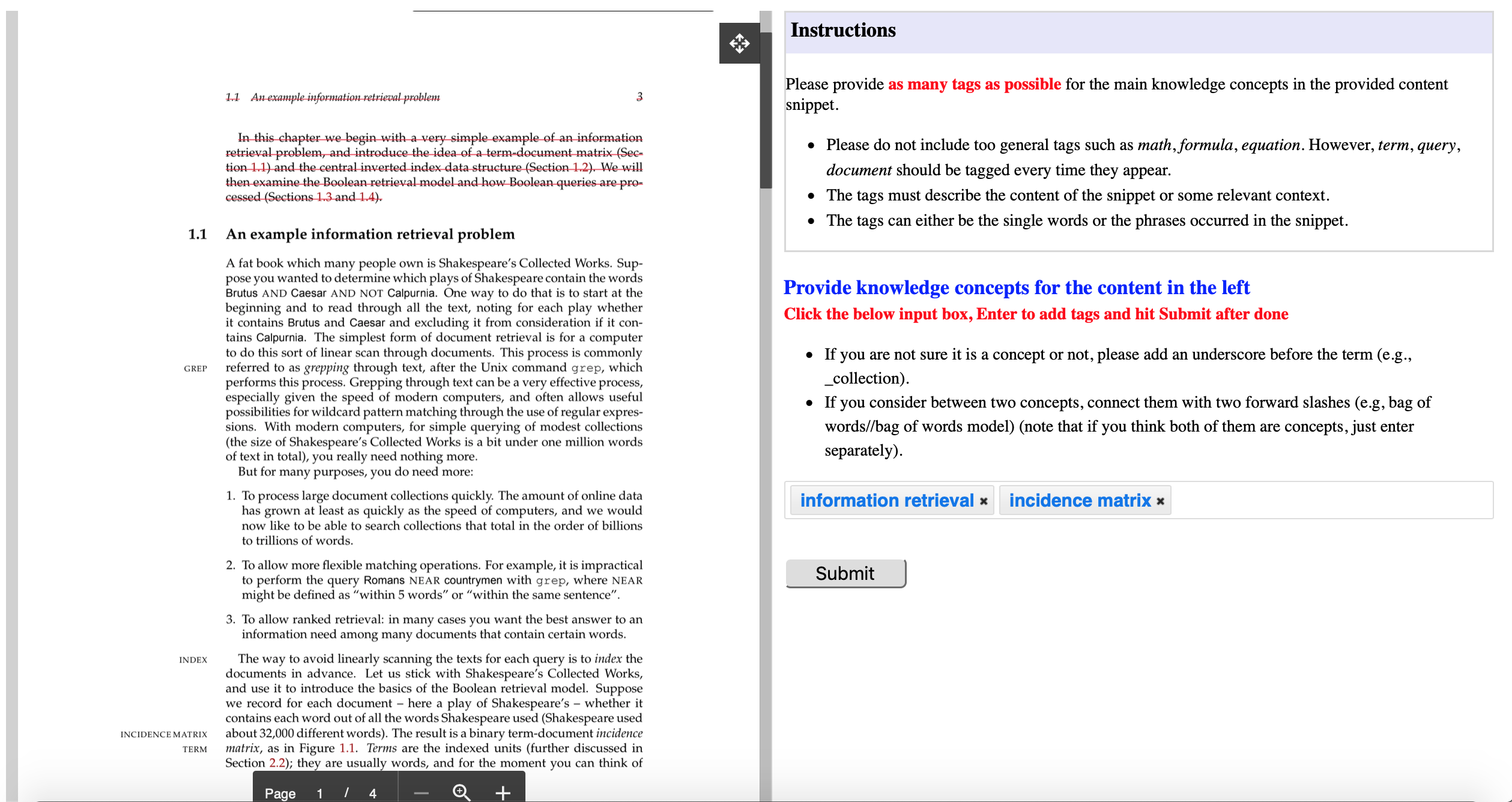}
    \caption{The main interface for annotating concepts.}
    \label{fig:interface}
\end{figure*}

To perform textbook annotation following the developed procedure, we hired three experts, one PhD student working in IR domain and two Master students who completed a graduate IR course with high final class scores. After eleven weeks, we replaced one Master student with a new Master student who also completed the IR course with high scores to see how the code book could help to achieve a good agreement rate with a new annotator. The PhD student was paid by the project and the three Master students were paid a stipend of \$12 per hour. The annotators were given task descriptions and the initial code book for annotating concepts (discussed in the next sections). Before staring the process, the annotators had to pass an annotation test and make themselves familiar with the task and the annotation interface (see Figure \ref{fig:interface}). 

\subsection{Task Description} Annotators were expected to work on one chapter per week for the first 16 chapters of IIR textbook (i.e., we only process these chapters because they are used in a real class room that students need to read them through an intelligent textbook interface). Each chapter includes multiple sections, which were considered as units or annotations. The sections were identified according to the headings in the table of content of the book (unless a section is too short and can be combined with the consecutive sections). The annotators were required to annotate all possible concepts which appear in the text of each section. Within a week, after completing annotating concepts, experts need to sit down together to discuss cases that they do not agree with one another, and come up with possible rules that help to increase the agreement.

\subsection{Initial Code Book}
\begin{figure*}
    \includegraphics[width = 1\textwidth]{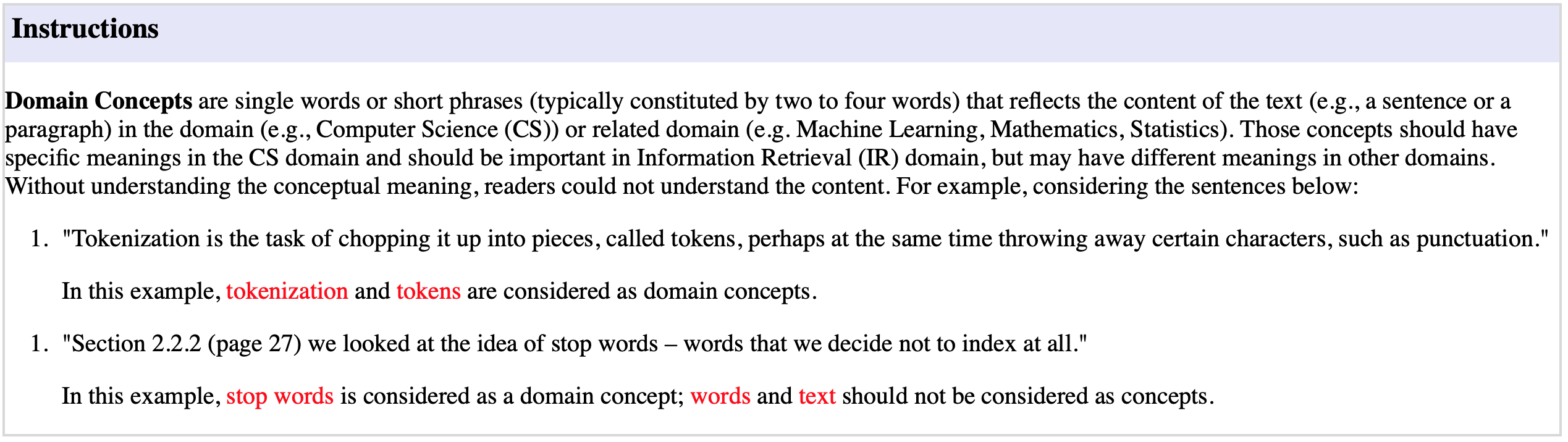}
    \caption{The initial code book for textbook concept annotation task.}
    \label{fig:instruction}
\end{figure*}

The annotators initially started performing the tasks by following a concept annotation instructions. The instructions shown to the annotators are depicted in Figure \ref{fig:instruction}. The instructions were developed by a group of experts in the field for the tagging tasks, and we consider it as the initial code book of our coding procedure. Throughout the coding process, the code book had to be updated and eventually become an outcome of the annotation procedure. 

\subsection{Annotating Process for the First Two Chapters}
The annotators started the annotation process following the procedures described above. They completed one chapter every week (called ``round'') via the annotation interface. At the beginning of each round the annotators tagged concepts section by section, which took  about 2-3 hours in total. The results (3 independent sets of annotations) were processed to identify agreement cases (i.e., the concepts tagged by all three experts) and disagreement cases (concepts that were tagged by two or only one expert). The annotators set up meetings to discuss disagreement cases they do not agree with one another and modify the results, which took another 2-3 hours. Based on the discussion and the analysis of disagreement cases, the code book was updated by adding or modifying the rules and the new agreement score was re-calculated after discussion. In the next round, annotators performed the annotation task based on the updated code book from the previous rounds. 

\subsection{Process Modification}
\label{process_modification}
After the first two rounds, we found out that the key reason of the low agreements before discussion is that the annotators unintentionally missed the concepts although they agree that these concepts should be tagged. To resolve this problem, we refined our annotation process by adding one more step: after completing their own annotation part, the experts were required to check missed concepts (see Figure \ref{process_modification}). It was done by reviewing a file where the experts could see each other's annotation results and decide whether they want to change their own annotations. The experts were asked to locate the missing concepts in the original context to make the decision. After checking the missing terms, the new agreement was calculated and the annotators discussed and updated the code book as described in the previous section.

\begin{figure}[htbp]
\centering
    \includegraphics[width = 0.5\textwidth]{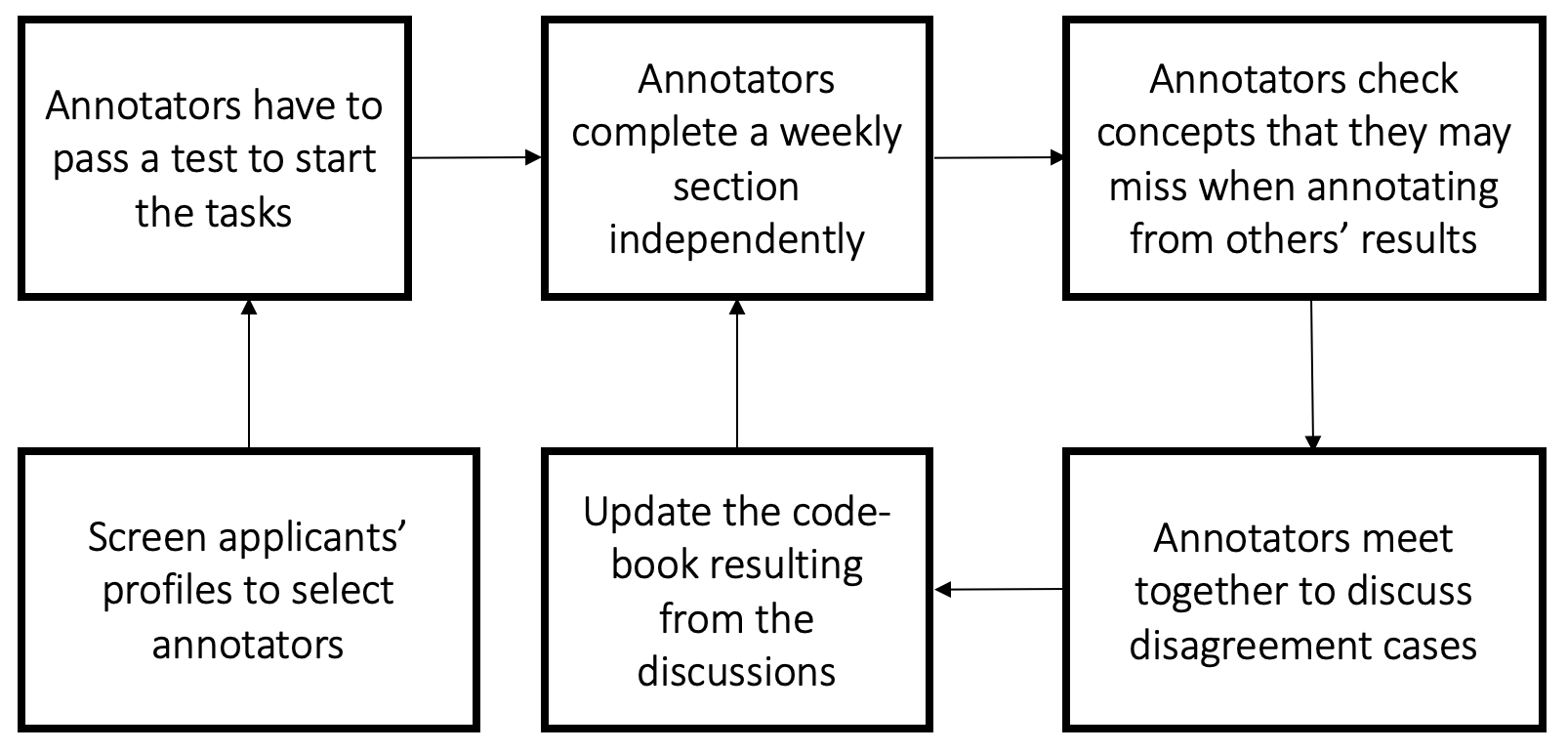}
    \caption{Modified Coding procedure diagram.}
    \label{fig:procedure_modified}
\vspace{-15pt}
\end{figure}

\subsection{Improvements from the Modified Process and Code Book}
To see the improvements after refining the coding procedure and to demonstrate the benefit of the incrementally improving code book, we are in process of working on reporting  inter-annotator agreement among the three annotators and also the average agreement of the pairs.





\section{The Outcomes}
In this section, we present the main outcomes of our attempts to develop a systematic concept annotation procedure. The outcomes include the final annotation procedure, the concept annotation code book and the Information Retrieval corpus including the text of the first 86 sections from the selected \textit{IIR} book and the list of concepts associated with each of the sections.

\subsection{Final Coding Procedure}
The final procedure for systematic concept annotation was developed in the process of full-scale practical testing of the initial procedure. While the initial procedure already integrated best practices reported in earlier publications, our thorough testing led to an important modification explained in the previous section. The final coding procedure shown in Figure \ref{fig:procedure_modified} includes the following steps:

\begin{itemize}
    \item \textbf{Step 1}: The project lead screen profiles of candidate annotators to choose annotators who satisfy specific criteria; for example: background knowledge.
    \item \textbf{Step 2}: The annotators make themselves familiar with the interface that is used to annotate knowledge components. The annotators also study the instructions that they need to follow in the annotation process. To ensure that they understood what they are asked to do and how to do it, the annotators had to pass a test related to the main tasks.
    \item \textbf{Step 3}: The annotators complete one round of annotations processing independently an assigned portion of text (in our case, one chapter every week) following the code book.
    \item \textbf{Step 4}: The annotators check potentially missed concepts by reviewing the annotation results produced by other annotators. They are required to locate the missing concepts in the original text to make decisions.
    \item \textbf{Step 5}: The annotators meet after finishing the annotation round to discuss disagreement cases and to come up with new rules to prevent the identified conflicts in the future.
    \item \textbf{Step 6}: The new rules from Step 5 are added to the code book (if necessary).
    \item \textbf{Step 7}: Switch to the next portion of text to be annotated and repeat the process starting from Step 3 until completing all text is annotated.
   
\end{itemize}
\subsection{Code book}


Table~\ref{tab:coding} lists the coding schema and detailed rules with examples of concepts and explanations. Following the coding procedure, we added one or more rules after each round. In total, we have ten rules. Most of the rules were added after the first few rounds (e.g., round 1,2,3). After round 9, no new rules were added. It indicates that the resulting table might be sufficiently complete and recommended for broader use. 

\begin{table*}[htbp]
\begin{centering}

\resizebox{\textwidth}{!}{\begin{tabular}{|l|l|l|}
\hline
\multicolumn{1}{|c|}{Rule} & \multicolumn{1}{c|}{Description} & \multicolumn{1}{c|}{Examples \& Explaination} \\ \hline
\begin{tabular}[c]{@{}l@{}}1.\\ (Round 1)\end{tabular} & Only noun/noun phrases are considered. & \begin{tabular}[c]{@{}l@{}}{\bf Concept:} \\ sorting algorithm, wildcard pattern matching, boolean retrieval model\\ {\bf Not concept:}\\ merging postings list, ranking documents \\\\

In the examples above, {\it merging postings list} and {\it ranking documents}\\ are not concepts, because they are not nouns or noun phrases.\end{tabular} \\ \hline
\begin{tabular}[c]{@{}l@{}}2.\\ (Round 1)\end{tabular} & Abbreviation of a concept is also a concept. & \begin{tabular}[c]{@{}l@{}}-IR (information retrieval)\\ -EM (expectation maximization)\\ \\{\it IR} and {\it EM} are all concepts, because {\it information retrieval} and \\ {\it expectation maximization} are concepts\end{tabular} \\ \hline
\begin{tabular}[c]{@{}l@{}}3.\\ (Round 1)\end{tabular} & \begin{tabular}[c]{@{}l@{}}Annotate the whole noun/noun phrases, \\ but ignore the general adj. (e.g., long, big etc.)\end{tabular} & \begin{tabular}[c]{@{}l@{}}{\bf Concept:}\\ latent linguistic structure, hidden variables \\ {\bf Not concept:}\\ long query, big document collection\\ \\ In the examples above, long and big are too general. \\ Only {\it query} and {\it document collection} are concepts.\end{tabular} \\ \hline
\begin{tabular}[c]{@{}l@{}}4.\\ (Round 2)\end{tabular} & \begin{tabular}[c]{@{}l@{}}If two noun phrases are concepts,\\ the combination should be the concept.\end{tabular} & \begin{tabular}[c]{@{}l@{}}{\bf Concept:} postings list data structure\\ \\ In the example above, {\it postings list} and {\it data structure} are concepts, \\ so {\it postings list data structure} is a concept.\end{tabular} \\ \hline
\begin{tabular}[c]{@{}l@{}}5.\\ (Round 3)\end{tabular} & \begin{tabular}[c]{@{}l@{}}The concepts combined with conjunctions \\ should be separated (e.g., and, or).\end{tabular} & \begin{tabular}[c]{@{}l@{}}- ``boolean and proximity queries"\\ \\ In the example above, you need to annotate the two\\ concepts {\it boolean queries} and {\it proximity queries}\end{tabular} \\ \hline
\begin{tabular}[c]{@{}l@{}}6.\\ (Round 5)\end{tabular} &  All variation of the concepts should be annotated. & \begin{tabular}[c]{@{}l@{}}-Multi-term query\\ -Bi-term query\\ -Three-term query\\ \\ The examples above are variation of the concept {\it query},\\ therefore they should be annotated.\end{tabular} \\ \hline
\begin{tabular}[c]{@{}l@{}}7.\\ (Round 6)\end{tabular} & \begin{tabular}[c]{@{}l@{}}Annotate all special / not general phrases\\  in computer science related domain\\ e.g., Statistics, mathematics\end{tabular} & \begin{tabular}[c]{@{}l@{}}{\bf Concepts:} quadratic function, binomial distribution\\ \\ {\it Quadratic function} and {\it binomial distribution} are  concepts, \\ because they are important phrases in Statistics domain.\end{tabular} \\ \hline
\begin{tabular}[c]{@{}l@{}}8.\\ (Round 6)\end{tabular} & Ignore the Abbreviation in brackets. & \begin{tabular}[c]{@{}l@{}}-inverse document frequency (idf)\\ -variable byte (vb)\\ -encodingmegabytes (mb)\\ \\ In the examples above, idf, vb and mb should be ignored\end{tabular} \\ \hline
\begin{tabular}[c]{@{}l@{}}9.\\ (Round 8)\end{tabular} & If the concept term has punctuations, keep them. & \begin{tabular}[c]{@{}l@{}}- (query, document) pairs\\ \\ The example above should be annotated as a concept\\ including the bracket and comma.\end{tabular} \\ \hline
\begin{tabular}[c]{@{}l@{}}10.\\ (Round 9)\end{tabular} & \begin{tabular}[c]{@{}l@{}}The well-known and important examples should\\be annotated. \end{tabular}& \begin{tabular}[c]{@{}l@{}}- A well-known example is the Unified Medical Language
System... \\ \\ {\it Unified Medical Language
System} should be annotated. \end{tabular} \\ \hline
\end{tabular}
}
\caption{Coding schema for concept annotation}
\label{tab:coding}
\end{centering}
\end{table*}


\subsection{The Corpus}

The important practical outcome of our work is the IR Corpus, which is the full set of annotations for the first 16 chapters (i.e., 86 sections) of \textit{Introduction to Information Retrieval} textbook. We make this data available on Github folder\footnote{https://github.com/PAWSLabUniversityOfPittsburgh/Concept-Extraction}, called SKA (i.e., Systematic Knowledge Annotation) corpus. Some process and outcome statistics for this corpus is shown in Table \ref{tab:dataset}. To stress the importance of the systematic annotation process, along with the data about final concepts (agreed by all the three experts after their discussions, see column 4\&5 in Table \ref{tab:dataset}), we also report the statistics for concepts that are annotated by all the experts before discussions (see column 2\&3 in Table \ref{tab:dataset}). Note that the number of concepts and unique concepts after discussions are larger than those before discussions. 

As also can be seen in Table \ref{tab:dataset}, the distribution of n-grams is very similar before and after discussion. For the final concept list, bi-grams contribute to about \textit{50\%} of all the concepts for both cases (i.e., \textit{number of concepts} and \textit{number of unique concepts}). The longer a concept is, the less frequent it in the corpus. Unique 1-grams account for \textit{18.02\%} of all the unique concepts while 1-grams alone account for \textit{35.31\%} of all the concepts. On the other hand, unique 3-grams account for \textit{21.39\%} of all the unique concepts while 3-grams only contribute to \textit{13.29\%} of all the concepts. This statistics could be helpful for designing automatic concept extraction; for instance, instead of trying to predict all the concepts, one just needs to focus on one to four grams which contribute to about \textit{99.5\%} to improve the model performance. 

\begin{table*}[htbp]
\begin{tabular}{|l|l|l|l|l|}
\hline
Characteristic     & \begin{tabular}[c]{@{}c@{}}Number of concepts \\ (before discussion)\end{tabular}   & \begin{tabular}[c]{@{}c@{}}Number of unique concepts \\ (before discussion)\end{tabular}   & \begin{tabular}[c]{@{}c@{}}Number of concepts \\ (after discussion)\end{tabular} & \begin{tabular}[c]{@{}c@{}}Number of unique concepts \\ (after discussion)\end{tabular} \\ \hline
1-grams & 958 (36.19\%)           & 236 (18.60\%)    & 1121 (35.31\%)           & 278 (18.02\%)                    \\ \hline
2-grams   & 1291 (48.77\%)               & 8719 (56.66\%)  & 1565 (49.29\%)               & 871 (56.45\%)                      \\ \hline
3-grams   & 351 (13.26\%)                & 270 (21.27\%)    & 422 (13.29\%)                & 330 (21.39\%)                      \\ \hline
4-grams   & 41   (1.55\%)             & 38     (2.99\%)  & 58    (1.83\%)             & 55     (3.56\%)                   \\ \hline
5+6-grams   & 6 (0.23\%)                 & 6    (0.47\%)   & 9 (0.28\%)                 & 9    (0.58\%)                     \\ \hline
all grams & 2647               & 1269        & 3175               & 1543                      \\ \hline
\end{tabular}
\caption{Data statistics of IR corpus. The concepts included in the final result are agreed by all the three experts before the discussions (i.e., column 1 \& 2) and after the discussions (i.e., column 3 \& 4).}
\label{tab:dataset}
\vspace{-1em}
\end{table*}


\section{Conclusion and Future work}
 In this paper, we present a reliable systematic knowledge engineering approach for fine-grained annotation of textbooks with underlying knowledge in the form of concepts. We explored this approach by performing a full-scale annotation procedure on a popular open source textbook \textit{Introduction to Information Retrieval} (IIR). In the process of working with IIR, we refined and finalized the proposed approach. This approach itself, the outcomes of our work include a code book, which can be used to annotate similar textbooks, and a public dataset. The dataset includes the textbook content and a full set of section-level annotation (SKA corpus) and could be used by the document engineering community to refine and evaluate their models. In  future we would like to  compared our SKA corpus against alternatively produced annotation corpora in terms of their performance on two target tasks performed by intelligent textbooks : document linking and student modeling. 
 
While the present work provides the first approach to annotate knowledge for intelligent textbooks, our work left a number of questions open. First, in this work, MTurk crowdworkers were asked to annotate without codebook. It remains to be seen whether the annotation produced by crowdworkers \emph{with} the codebook could reach the quality of the experts. Second, the concepts extracted by IBM automatic approach produce good results in both tasks, which encourages us to explore a hybrid approach which combines the automatic extraction method and the systematic procedure. The automatic extraction method may have potential power of improving the quality of the annotation as well as reducing the annotation load. 

\bibliographystyle{unsrt}  
\bibliography{template}

\end{document}